# Crystalline Water Structure in Room-Temperature Clathrate State: Hydrogen-Bonded Pentagonal Rings

Ching-Hsiu Chen[1], Wei-Hao Hsu[1], Ryoko Oishi-Tomiyasu[2], Chi-Cheng Lee[3], Ming-Wen Chu[4], and Ing-Shouh Hwang[1]*

[1] Institute of Physics, Academia Sinica, Nankang, Taipei 115, Taiwan

[2] Institute of Mathematics for Industry, Kyushu University, Fukuoka, 819-0395, Japan

[3] Department of Physics, Tamkang University, New Taipei 251301, Taiwan

[4] Center for Condensed Matter Sciences, National Taiwan University, Taipei 10617, Taiwan

*Corresponding author. ishwang@phys.sinica.edu.tw (I.-S. Hwang)

## Abstract

Water hydrogen bonding is extremely versatile; approximately 20 ice structures and several types of clathrate hydrate structures have been identified. These crystalline water structures form at temperatures below room temperature and/or at high pressure. We used transmission electron microscopy to study a new crystalline water structure in a clathrate state that is prepared by sandwiching gas-supersaturated water between graphene layers under ambient conditions. In this clathrate state, water molecules form a three-dimensional hydrogen bonding network that encloses gas-filled cages 2–4 nm in size. We derived the crystalline water structure by recording and analyzing electron diffraction patterns and performing first-principles calculations. The structure consists purely of pentagonal rings and has a topology similar to that of water ice XVII. The study proposed a mechanism for the formation of the clathrate state. The present results improve the understanding of interactions among water and small nonpolar molecules and offer novel insights into the local structures of ambient liquid water.

Water is one of the most important substances on Earth because it plays vital roles in various biological, chemical, physical, geological, and technological processes. It has many unique properties compared to other liquids, primarily because of its ability to participate in extensive hydrogen bonding interactions. Water is arguably the most important solvent in nature because of its ability to dissolve a wide variety of substances. It serves as an excellent solvent for ionic and polar solutes. Additionally, water can dissolve a limited number of small hydrophobic (nonpolar) molecules, but how water molecules reorganize around nonpolar molecules, hydrophobic hydration, under ambient conditions remains highly debated. The dissolution of simple gases or hydrocarbons (small nonpolar molecules) in water exhibits abnormal thermodynamic properties; for example, this dissolution can cause an anomalously large increase in heat capacity, a large decrease in standard enthalpy, and an extremely large loss of entropy[1,2]. In 1945, Frank and Evans proposed the "iceberg model"[3], in which water forms frozen patches or ice-like structures around individual nonpolar molecules, to explain the abnormal thermodynamic properties. Numerous experimental and theoretical investigations of the structuring of water (hydration) around small nonpolar solute molecules have been conducted. Some studies have obtained evidence supporting enhanced water ordering or increased tetrahedral order around small hydrophobic groups in aqueous solutions [4–8], but other works do not support the "iceberg" view of hydrophobic hydration[1,9–13].

Due to the low solubility and concentration of small nonpolar molecules in liquid water, hydrophobic molecules have generally been assumed to be well dispersed as monomers, and the abnormal thermodynamic properties have been presumed to result from the hydration of individual hydrophobic molecules. Recent results obtained through transmission electron microscopy (TEM) of gas-supersaturated water encapsulated in graphene liquid cells (GLCs) have challenged these assumptions; the TEM study revealed the presence of a special type of mesoscopic clathrate hydrate structures—one that has a high number density of tiny cells that often exhibit a honeycomb-like morphology[14]. A few discrete sets of strong diffraction spots corresponding to certain crystal planes were observed. The major constituent of the clathrate structures is water, and the d-spacings (interplanar spacings) of the crystalline structures are not different for different gases[14], suggesting that water molecules form the crystalline structures that produce diffraction spots. Dark-field TEM imaging of the clathrate structures has indicated that water molecules form crystalline structures surrounding or between gas-containing cavities. This novel clathrate state differs from typical states of gas clathrates (gas hydrates) in several aspects: (1) it exists at room temperature (RT), (2) the cages are larger (typically 2–4 nm), (3) the honeycomb-like cell morphology has neither long-ranged translational nor orientational order, and (4) the state has a lateral size smaller than 1 μm. The observation of these mesoscopic clathrate hydrate structures indicates that small nonpolar molecules may not be homogeneously dispersed as monomers in liquid water and that the abnormal thermodynamic properties may result from the formation of the clathrate hydrate structures rather than being related to the hydration of individual hydrophobic solutes.

Another TEM study of ethanol-water (EW) mixtures (~10% volume fraction ethanol) encapsulated in GLCs under ambient conditions also reported mesoscopic clathrate structures resembling those in gas-supersaturated water[15]. This is likely due to the mixing of ethanol with water causing gas supersaturation, as air gas is more soluble in the individual components than in the mixture[16]. The presence of the mesoscopic clathrate structures explains many long-standing puzzles related to EW mixtures, such as a sharp decrease in excess entropy and enthalpy[17], a sharp increase in excess heat capacity[18], a sharp increase in the strength of hydrogen bonds[19–21] with increasing ethanol concentration in the low ethanol concentration regime, and the formation of bulk "nanobubbles" upon the mixing of water with ethanol.[15]

The crystalline water structure in this novel clathrate state has a d-spacing of 4.5–4.8 Å[14,15], which has not been reported for any other water ice or clathrate hydrate structure. Determining this new crystalline water structure is thus of considerable scientific interest. Because of the small size and low concentration of the mesoscopic clathrate hydrate structures, techniques such as X-ray diffraction and neutron diffraction, which are predominantly used to determine crystalline water structures, are not applicable. In the present study, TEM was adopted because of its ability to image structures at a nanometer or subnanometer resolution and to record selected area electron diffraction (SAED) patterns for regions as small as several tens of nanometers. We derived the crystalline water structure of the mesoscopic clathrate structures by recording and analyzing in-zone electron diffraction patterns (IZEDPs) and by performing first-principles calculations. The IZEDP evidence confirmed that the clathrate structures that formed in gas-supersaturated water and in EW mixtures have the same crystalline water structure Determining the water structure is essential for understanding the stability and formation mechanism of this clathrate state as well as the interactions between water and small non-polar molecules. It also helps further theoretical and experimental characterization of various properties of this new clathrate hydrate state.

**Results and Discussion**

The sample preparation and TEM have been detailed in previous publications[14,15] and the Experimental Section. All experiments were conducted at RT (22–24 °C) unless otherwise specified. Fig. S1 A,B presents bright-field TEM images of a mesoscopic clathrate structure sandwiched in a GLC. A high number density of nanometer-scale white dots was observed at underfocus (Fig. S1A) and the white dots became dark spots at overfocus (Fig. S1B); little contrast was seen in focus (data not shown)[14,15]. Fig. S1C presents an SAED pattern of the clathrate structure. Fig. S1D displays a dark-field image acquired from a diffraction spot within the pattern in Fig. S1C and reveals honeycomb-like patterns with bright features surrounding the tiny gas-containing cells. The morphology and contrast behavior of this study's bright-field and dark-field images are identical to those reported previously[14,15]. Many of the mesoscopic clathrate structures exhibited a single crystal grain. We tilted the samples to acquire IZEDPs for each clathrate structure; more than one IZEDPs could often be acquired in the range of our TEM goniometer (x, y ± 30°). Fig. 1A–D presents a set of IZEDPs acquired for a clathrate structure (Clathrate 1) in 10% EW mixture (IZEDPs categorized as Types $\alpha$–$\delta$, respectively). Types $\alpha$–$\gamma$ are three IZEDPs that were often observed for clathrate structures in EW mixtures as well as in gas-supersaturated water; Type $\alpha$ (Fig. 1A) was the most frequently observed IZEDP for different clathrate structures. The relative orientations among the IZEDPs of Clathrate 1 were calculated on the basis of the recorded tilt angles and are presented in Table 1.

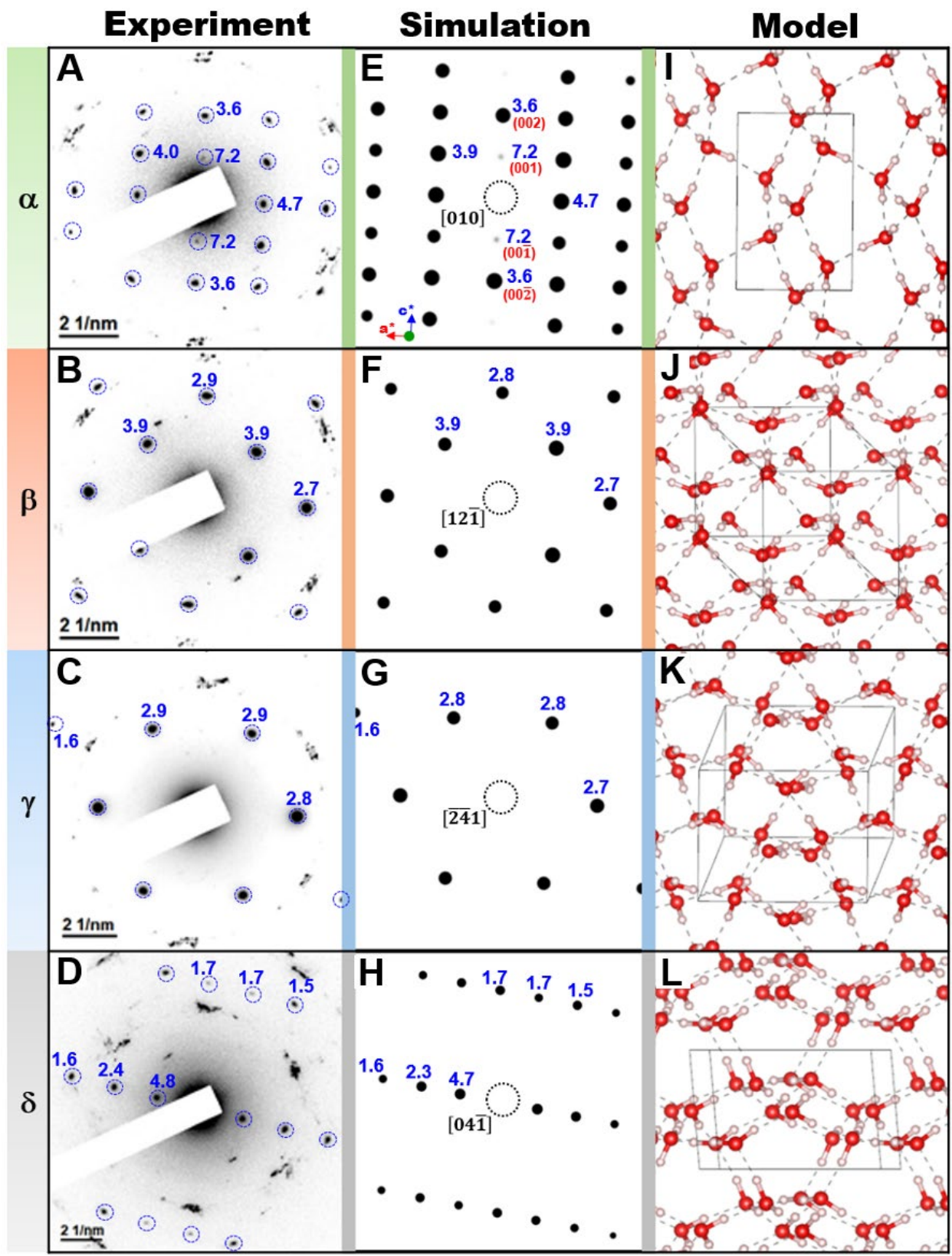

**Figure 1.** Four types of IZEDP acquired for Clathrate 1 and the corresponding simulated diffraction patterns and models. (A–D) IZEDPs, categorized as Types **α–δ**, respectively, acquired by tilting the sample with the TEM goniometer. The sample angles are (x = +31.7°, y = +16.0°) for Type **α**, (x = +2.0°, y = −19.0°) for Type **β**, (x = +0.7°, y = −0.3°) for Type **γ**, and (x = +31.0°, y = −2.0°) for Type **δ**. The relative angles between these four sample orientations were calculated and are listed in Table 1. The diffraction spots associated with the clathrate structure are circled. The numbers next to some diffraction spots indicate the d-spacing (unit: Å). The contrast of the patterns is reversed for better comparison with simulated diffraction patterns. (E–H) Simulated (dynamic) electron diffraction patterns based on the CS-RT model for Types **α–δ**, respectively. The zone axis for each diffraction pattern is indicated next to the central spot. The number next to some diffraction spots indicates the d-spacing (unit: Å). (I–L) View of calculated water structures in the [010] (*b*-axis), [12$\bar{1}$], [$\bar{2}$4$\bar{1}$], and [04$\bar{1}$] directions, respectively.

|  | Clathrate 1 | | | Clathrate 2 | | Simulation | | |
| --- | --- | --- | --- | --- | --- | --- | --- | --- |
|  | Type-α | Type-β | Type-γ | Type-α | Type-β | Type-α | Type-β | Type-γ |
| Type-β | 44.6° | --- | --- | 44.7° | --- | 46.4° | --- | --- |
| Type-γ | 34.6° | 18.7° | --- | 34.1° | 16.6° | 36.0° | 16.4° | --- |
| Type-δ | 18.0° | 33.2° | 30.3° | ------ | | 18.2° | 32.4° | 28.4° |

Table 1 Experimental relative angles between various IZEDP types and simulated angles based on the CS-RT structure. Experimental angles were measured on two clathrate structures (Clathrate 1 and Clathrate 2). The angular uncertainty of our goniometer was approximately ±2°.

We derived the lattice parameters of the structure and indexed the IZEDPs by using the CONOGRAPH software (Supplementary Note 1); a hexagonal lattice with *a* = *b* = 5.40 Å and *c* = 7.12 Å was determined. First-principles calculations were performed to locate the equilibrium positions of atoms. Six water molecules were placed inside the unit cell because the mass density (0.99 g/cm$^3$) is close to that of liquid water. Fig. 2A shows the structure derived from the first-principles calculation, referred to as CS-RT. The full crystallographic information is displayed in

Table 2. In the obtained crystal structure, all water molecules are hydrogen-bonded to four others, and some distortions from the perfect tetrahedral structure are evident. The structure comprises only five-membered (pentagonal) rings of hydrogen-bonded water molecules, in contrast to the six-membered rings in hexagonal ice. The pentagonal rings are distorted and slightly puckered, differing from the planar pentagons in the typical clathrate hydrates CS-I, CS-II, and CS-III. The space group of the structure is P1 (triclinic; Supplementary Note 2). The structure viewed along the *a*- or *b*-axis is presented in Fig. 2B. The spiraling channels formed by water molecules can be identified along the *c*-axis (Fig. 2C). The topology of oxygen atoms in this crystal structure is identical to that of ice XVII[22], but the lattice parameters are different. The O–O distances are shorter than those in hexagonal ice ($d_{OO} \approx 2.75$ Å), indicating stronger hydrogen bonds in the mesoscopic clathrate structure.

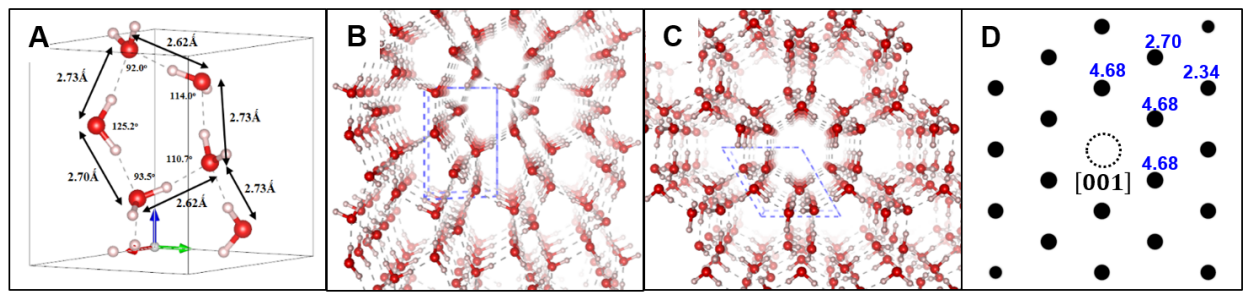

**Figure 2. Crystalline water structure in CS-RT, derived from first-principles calculations.** (A) Unit cell of the structure with *a* = 5.40 Å, *b* = 5.40 Å, *c* = 7.12 Å, *α* = 90°, *β* = 90°, and *γ* = 120°. (B, C) View of the water structure along the *a*- and *c*-axes, respectively. (D) Simulated IZEDP along the *c*-axis. The CS-RT structure is derived from diffraction data, and only spatially averaged atomic positions are obtained. A certain degree of disorder in hydrogen bonding can be expected in the mesoscopic clathrate structures.

| Atom | x | y | z |
|------|-------|-------|-------|
| O1 | 0.712 | 0.453 | 0.301 |
| O2 | 0.479 | 0.727 | 0.457 |
| O3 | 0.488 | 0.225 | 0.957 |
| O4 | 0.754 | 0.227 | 0.630 |
| O5 | 0.214 | 0.725 | 0.130 |
| O6 | 0.256 | 0.499 | 0.801 |
| H1 | 0.907 | 0.576 | 0.246 |
| H2 | 0.643 | 0.582 | 0.352 |
| H3 | 0.225 | 0.613 | 0.022 |
| H4 | 0.061 | 0.376 | 0.747 |
| H5 | 0.330 | 0.722 | 0.238 |
| H6 | 0.325 | 0.370 | 0.852 |
| H7 | 0.584 | 0.336 | 0.073 |
| H8 | 0.638 | 0.229 | 0.738 |
| H9 | 0.598 | 0.923 | 0.513 |
| H10 | 0.369 | 0.029 | 0.013 |
| H11 | 0.742 | 0.339 | 0.522 |
| H12 | 0.384 | 0.616 | 0.574 |

Table 2 Fractional atomic coordinates of CS-RT derived from the first-principles calculation. The space group of the structure is P1. The lattice constants are $a = b = 5.40$ Å and $c = 7.12$ Å.

The four types of IZEDP presented in Fig. 1A–D were indexed as being along [010] for Type **α**, [12$\bar{1}$] for Type **β**, [$\bar{2}4$1] for Type **γ**, and [04$\bar{1}$] for Type **δ**. Fig. 1E–H shows the simulated IZEDPs for Types **α**–**δ**, respectively; the related angles and d-spacings of the simulated patterns agree well with those of the experimental IZEDPs (Fig. S2). Fig. 1I–L presents the corresponding structure for the four types of IZEDP. The d-spacings of the same types of IZEDP were discovered to vary slightly from clathrate to clathrate (Fig. S3). We also observed a slow change in the d-spacings of IZEDPs for the same clathrate over time, despite the IZEDPs remaining roughly the same. The variation in the d-spacings for a given IZEDP type was 3%–5%, with this variation potentially being due to the substantial thermal motion of water molecules at RT. Considerable thermal motion would also lead to a gradual change in the cell morphology of the clathrate structures over time[14], indicating that even when long-range order is maintained, hydrogen-bonded networks are continually disrupted and formed.

We note that the reflections 00$l$, where $l$ is an odd number, are very weak in the simulated dynamic diffraction pattern of Type **α** (Fig. 1E) ; this finding is consistent with the experimental findings regarding the IZEDPs of Type **α** (Figs. 1A and S3A1–3). These spots are actually caused by multiple scattering, as indicated by their disappearance when the sample was rotated around the *c*-axis by several degrees (Fig. S4A,B). The kinematic diffraction of Type **α** that was simulated on the basis of the CS-RT model also exhibited disappearance of these diffraction spots, indicating that reflections 00$l$, where $l$ is an odd number, are kinematically forbidden (Fig. S4C). In particular, the diffraction spots with d-spacing of 6.8–7.2 Å (Figs. 1A and S3A1–3), corresponding to reflections 001 and 00$\bar{1}$, were observed only in Type **α** IZEDPs. The reflections 00$l$, where $l$ is an even number, were allowed, which is consistent with our observation that reflections 002 and 00$\bar{2}$ with d-spacing of 3.4–3.6 Å were often observed in electron diffraction patterns recorded for clathrate structures[14,15].

Fig. S5 presents three IZEDPs recorded for another clathrate structure in $N_2$-supersaturated water (Clathrate 2). Types **α**–**γ** were obtained, and the relative orientations between the three types of IZEDP were close to those for Clathrate 1 (Table 1). We observed Types **α**–**δ** for many mesoscopic clathrate structures formed in gas-supersaturated water as well as in EW mixtures; this indicates that the clathrate structures formed under these two conditions have the same crystalline water structure. In another clathrate structure (Clathrate 3), we recorded IZEDPs of Types **α** and **γ** and a different type of IZEDP (Type **ε**; Fig. S6). Type **ε** was indexed as along [211]. Table S1 lists the experimental and simulated angles between the three IZEDPs. The favorable agreement between the experimental and simulated IZEDPs and the relative inter-IZEDP angles (Tables 1 and S1) in several clathrate structures indicate the validity of the CS-RT model.

The IZEDP along the *c*-axis should exhibit diffraction spots of nearly six-fold symmetry (Fig. 2D), but we never obtained a diffraction pattern with such spots. The *c*-axis of the crystals was likely to have been roughly parallel to the graphene layer (nearly perpendicular to the electron beam or thickness direction of the GLC). This is likely what occurred, given that Type *α* was the most frequently observed IZEDP and that reflections $00\bar{2}$ and $00\bar{2}$ with d-spacings of 3.4–3.6 Å were often discovered in electron diffraction patterns (Supplementary Note 3). This preferred crystal orientation is probably due to the relatively flat shape of the clathrate structures (thickness of ≈1/10 of the lateral size) in GLCs. Extending the CS-RT structure along the *c*-axis may be energetically favorable.

The mesoscopic clathrate structures observed in this study may have only existed in GLCs (Supplementary Note 4), but several experimental results suggested the presence of the mesoscopic clathrate structures in bulk water as well as in aqueous solutions. In water supersaturated with krypton or xenon, dynamic light scattering measurements indicated the existence of colloids with a size of tens to hundreds of nanometers; the colloids were suggested to be clathrate hydrate nanostructures because their mass density was determined to be very close to that of liquid water[23]. The gas-supersaturated water is prepared through equilibration of water with high-pressure gas for hours or longer and subsequent depressurization to ambient pressure; the method was similar to the method employed for our preparation of gas-supersaturated water. In another study of liquid water in equilibration with high-pressure gas (Kr, Xe, $CH_4$, and $C_2H_6$) at RT, infrared spectra revealed the existence of structures in which water hydrogen bonds are strengthened to levels observed in ice and clathrates[8]. We hypothesize that mesoscopic clathrate hydrate structures form in water in equilibration with gas above a certain pressure (~1 atm) and that the concentration and/or size of the mesoscopic clathrate structures increases with increasing pressure; the clathrate structures remain metastable when the water is depressurized from high to ambient pressure. This process would explain our measurements and TEM observation of mesoscopic clathrate structures. Similar processes have previously been discovered in EW mixtures. Objects with a size of tens to hundreds of nanometers have been identified in EW mixtures[15,21]; these nano-objects are generally believed to contain air gas because ethanol dissolves much more gas than water does and the gas supersaturation level increases with an increasing ethanol percentage in a low-ethanol-concentration regime[16]. A refractive index (1.27 ± 0.02) close to that of water ice was previously measured through nanoparticle tracking analysis of nano-objects in 10% EW mixtures[15]. Vibrational spectroscopies, such as Raman scattering and IR absorption techniques, have indicated that the strength of hydrogen bonds increases sharply with an increasing ethanol concentration in a low-ethanol-concentration regime[19–21], and the formation of clathrate-like structures in EW mixtures has thus been proposed[19,20]. These experimental results are consistent with our TEM finding of mesoscopic clathrate hydrate structures in 10% EW mixtures. Additionally, our earlier TEM study showed that liquid water and mesoscopic clathrate structures coexisted in the same water pocket of a GLC[14]. A recent atomic force microscopy (AFM) study of ozone nanobubbles adsorbed on a positively charged hydrophilic substrate indicated many semispherical nano-objects (several tens of nanometers in diameter) with tiny protrusions of the size of a few nanometers covering the entire surfaces of the nano-objects, revealing the surface to be solid[24]. The semispherical nano-objects (nanobubbles) may be the mesoscopic clathrate structures reported herein because the size of each surface protrusion observed through AFM is close to that of a cell in the mesoscopic clathrate structures of the present study.

The structure and dynamics of liquid water near RT under ambient pressure remain highly debated[25–28]. Some researchers have proposed the occurring of pentagonal rings in the fluctuating hydrogen-bonded network of liquid water[29–36]. The structure presented herein, CS-RT, supports this hypothesis. Pentagonal rings in liquid water may act as a source of frustration—that is, they may prevent liquid water from crystallizing into hexagonal or cubic ice. This could explain the considerable degree of supercooling observed in water before homogenous crystal nucleation occurs[35]. Pentagonal rings of hydrogen-bonded water molecules have also been discovered inside crystals of biological molecules under ambient temperature and pressure. The crystal structure of an alanine-rich antifreeze protein contains approximately 400 water molecules, which mainly form pentagonal rings with occasional tetragonal and hexagonal rings at the protein's core[7]. Additionally, one study reported a cluster of pentagonal arrays made up of 16 water molecules at a hydrophobic, intermolecular cleft in crambin crystals[38]. These findings along with the current discovery of a CS-RT structure strongly suggest the existence of local structures with pentagonal rings of hydrogen-bonded water molecules in ambient water or ambient aqueous solutions and indicate that the water structures with pentagonal rings may be stabilized when they are interfaced with hydrophobic surfaces with a large curvature (radius of curvature ≈ 1–2 nm).

We propose a scenario in which mesoscopic clathrate hydrate structures can form. When the gas concentration in water is far below the saturation level near RT and ambient pressure, nearly all gas atoms or molecules are well dispersed as monomers. When the gas concentration is near the saturation level, some of the dissolved gas monomers aggregate into clusters. A gas concentration above saturation results in the formation of more gas clusters. Local structures with pentagonal rings might form momentarily in liquid water and become stable when they form next to gas clusters of a size of 2–4 nm. Further formation of structured water comprising mainly pentagonal rings surrounding the gas cluster leads to the formation of a cell (Fig. 3A). Isolated cells undergo Brownian motion in liquid water and gradually aggregate as they encounter each other. The cell aggregates continue to grow and eventually reach a mesoscopic size (tens to hundreds of nanometers). In the current study, isolated cells were observed in TEM images of gas-supersaturated water in GLCs (Fig. 3 B,C), and small aggregates of cells have been observed in this study and previously (Fig. 3 B,C, and Fig. 9 in Ref. [14]). Once the cells have aggregated, the surrounding water molecules and those between the gas-containing cavities rearrange themselves into a CS-RT crystal with long-range order, which can be detected using SAED.

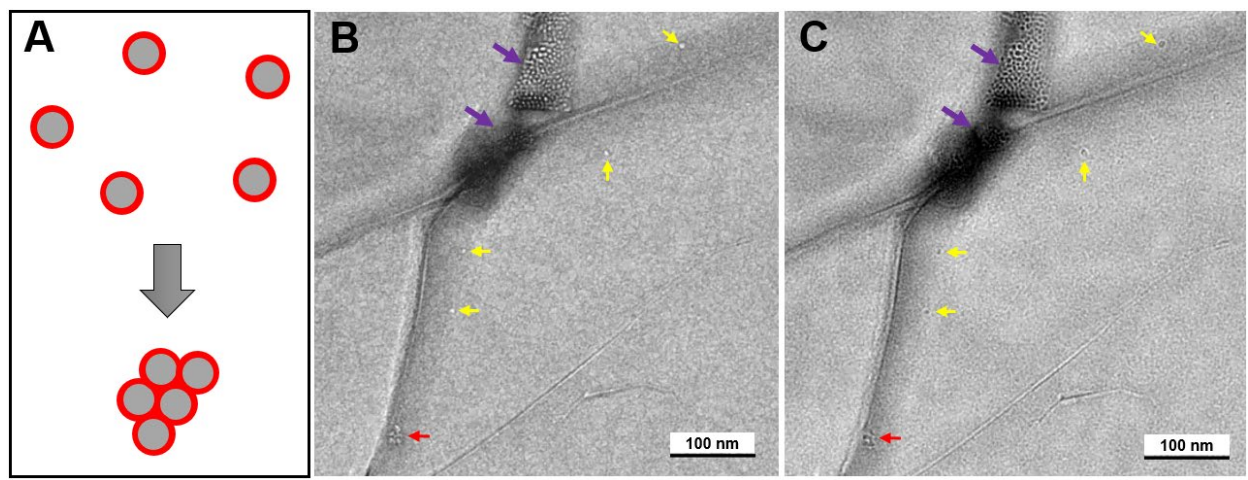

Fig. 3 Mechanism of formation of mesoscopic clathrate hydrate structures. (A) Formation of isolated cells with a solid-like water shell (indicated in red) surrounding a gas cluster, and aggregation of the cells. The solid-like water shell contains mainly pentagonal rings of hydrogen-bonded water molecules. (B, C) Bright-field TEM images of clathrate structures (purple arrows), isolated cells (yellow arrows), and a small cell aggregate (red arrows) acquired at underfocus and overfocus, respectively. The investigated sample is $N_2$-supersaturated water encapsulated in a GLC.

In typical gas hydrates, cavities are small (<1 nm) and are parts of the hydrate's periodic lattice. By contrast, the cavities in mesoscopic clathrate structures are considerably larger and not parts of the crystalline structure; it is the water between and surrounding the cavities that forms the crystalline structure (i.e., CS-RT). Further study is needed to understand how the cavities and the gas molecules inside the cavities stabilize the CS-RT structure and why the clathrate hydrate state appears to have a size limitation. Further experimental and theoretical investigations of mesoscopic clathrate structures would considerably improve the understanding of water's behaviors under ambient conditions, which are of substantial relevance to numerous physical, chemical, and biological systems in water. Such an understanding has potential to resolve many mysteries regarding ambient water as well as gas in water, such as hydrophobic hydration, nanobubbles in bulk water, and the disputed Mpemba effect[39,40].

**Experimental**

**Materials and Sample Preparation.** The preparation of gas-supersaturated water and EW mixtures (~10% volume fraction ethanol) and that of GLCs have been described in previous articles[14,15]. In the present study, two types of gas-supersaturated water (supersaturated with $N_2$ and $CO_2$) were prepared. Fig. S7 illustrates the procedures used to prepare GLCs and for TEM characterization.

**TEM.** All GLC samples were imaged using a field-emission TEM (JEM-2100F, JEOL) and Gatan imaging camera. To prevent sample damage, the electron acceleration voltage was set to 100 kV, and the emission current was approximately 182 µA. The pressure in the sample chamber was approximately 2 × $10^{-5}$ Pa. The TEM goniometer was used to tilt the sample within the range ±30° (x, y) to obtain diffraction patterns for different zone axes and to adjust the beam focus to ensure it aligned with the image position during the tilting process. Given the sensitivity of clathrate structures to high electron flux, we generally imaged the structures at magnifications lower than ×30,000 (dose rate ≈ 4.5 $e^{-}Å^{-2}s^{-1}$) to minimize the effect of electron irradiation[14].

**Indexing the IZEDPs and determination of the unit cell.** The lattice parameters of the unit cell were derived from the TEM diffraction patterns by using the CONOGRAPH[41] software. The unit-cell determination is based on the powder-indexing process, wherein reflection indices (***hkl***) were assigned to diffraction peaks. Once a sufficient number of reflection indices have been assigned to specific peaks, the unit-cell parameters could be determined. The details of the algorithm are presented in Supplementary Note 1.

**Structures derived from first-principles calculations.** First-principles calculations based on density functional theory were performed using the OpenMX code within the generalized gradient approximation[42] with a long-range dispersion correction (DFT-D2)[43]. Norm-conserving pseudopotentials and optimized pseudoatomic basis functions were adopted[44]. Two optimized radial functions were allocated for the s orbital and one for the p orbital for each H atom with a cutoff radius of 7 bohr, denoted H7.0-s2p1; O7.0-s2p2d1 was adopted for each O atom. A cutoff energy of 500 Ry was selected for numerical integrations and solution of the Poisson equation. The experimental lattice constants a = 5.40 Å and c = 7.12 Å were adopted for the calculations, which were performed with a 5 × 5 × 4 k-point mesh. Six water molecules were relaxed until all the atomic forces were smaller than 0.0001 hartrees/bohr. The VESTA software was used to draw atomic model figures[45].

**Simulation of electron diffraction.** Simulation of electron diffraction patterns was performed using the free software ReciPro[46], and the results were double-checked using the open-source Diffsims python library. The parameters in the TEM experiments were used in the simulations. The thickness of the sample in the simulation was set to approximately 30–50 nm.

[References]

1. Blokzijl, W. & Engberts, J. B. F. N. Hydrophobic Effects. Opinions and Facts. *Angew. Chem. Int. Ed. Engl.* **32**, 1545–1579 (1993).
2. Wilhelm, Emmerich., Battino, Rubin. & Wilcock, R. J. Low-pressure solubility of gases in liquid water. *Chem. Rev.* **77**, 219–262 (1977).
3. Frank, H. S. & Evans, M. W. Free Volume and Entropy in Condensed Systems III. Entropy in Binary Liquid Mixtures; Partial Molal Entropy in Dilute Solutions; Structure and Thermodynamics in Aqueous Electrolytes. *The Journal of Chemical Physics* **13**, 507–532 (1945).
4. Haselmeier, R., Holz, M., Marbach, W. & Weingaertner, H. Water Dynamics near a Dissolved Noble Gas. First Direct Experimental Evidence for a Retardation Effect. *J. Phys. Chem.* **99**, 2243–2246 (1995).
5. Rezus, Y. L. A. & Bakker, H. J. Observation of Immobilized Water Molecules around Hydrophobic Groups. *Phys. Rev. Lett.* **99**, 148301 (2007).
6. Davis, J. G., Gierszal, K. P., Wang, P. & Ben-Amotz, D. Water structural transformation at molecular hydrophobic interfaces. *Nature* **491**, 582–585 (2012).
7. Galamba, N. Water's Structure around Hydrophobic Solutes and the Iceberg Model. *J. Phys. Chem. B* **117**, 2153–2159 (2013).
8. Grdadolnik, J., Merzel, F. & Avbelj, F. Origin of hydrophobicity and enhanced water hydrogen bond strength near purely hydrophobic solutes. *Proc. Natl. Acad. Sci. U.S.A.* **114**, 322–327 (2017).
9. Turner, J., Soper, A. K. & Finney, J. L. A neutron-diffraction study of tetramethylammonium chloride in aqueous solution. *Molecular Physics* **70**, 679–700 (1990).
10. Qvist, J. & Halle, B. Thermal Signature of Hydrophobic Hydration Dynamics. *J. Am. Chem. Soc.* **130**, 10345–10353 (2008).
11. Koh, C. A., Wisbey, R. P., Wu, X., Westacott, R. E. & Soper, A. K. Water ordering around methane during hydrate formation. *The Journal of Chemical Physics* **113**, 6390–6397 (2000).
12. Buchanan, P., Aldiwan, N., Soper, A. K., Creek, J. L. & Koh, C. A. Decreased structure on dissolving methane in water. *Chemical Physics Letters* **415**, 89–93 (2005).
13. Bowron, D. T., Filipponi, A., Lobban, C. & Finney, J. L. Temperature-induced disordering of the hydrophobic hydration shell of Kr and Xe. *Chemical Physics Letters* **293**, 33–37 (1998).
14. Hsu, W.-H. & Hwang, I.-S. Investigating states of gas in water encapsulated between graphene layers. *Chem. Sci.* **12**, 2635–2645 (2021).

15. Hsu, W.-H. *et al.* Observation of mesoscopic clathrate structures in ethanol-water mixtures. *Journal of Molecular Liquids* **366**, 120299 (2022).
16. An, H., Liu, G., Atkin, R. & Craig, V. S. J. Surface Nanobubbles in Nonaqueous Media: Looking for Nanobubbles in DMSO, Formamide, Propylene Carbonate, Ethylammonium Nitrate, and Propylammonium Nitrate. *ACS Nano* **9**, 7596–7607 (2015).
17. Franks, F. & Ives, D. J. G. The structural properties of alcohol–water mixtures. *Q. Rev. Chem. Soc.* **20**, 1–44 (1966).
18. Benson, G. C. & D'Arcy, P. J. Excess isobaric heat capacities of water-n-alcohol mixtures. *J. Chem. Eng. Data* **27**, 439–442 (1982).
19. Dolenko, T. A. *et al.* Raman Spectroscopy of Water–Ethanol Solutions: The Estimation of Hydrogen Bonding Energy and the Appearance of Clathrate-like Structures in Solutions. *J. Phys. Chem. A* **119**, 10806–10815 (2015).
20. Burikov, S., Dolenko, T., Patsaeva, S., Starokurov, Y. & Yuzhakov, V. Raman and IR spectroscopy research on hydrogen bonding in water-ethanol systems. *Mol. Phys.* **108**, 2427–2436 (2010).
21. Jadhav, A. J. & Barigou, M. Bulk Nanobubbles or Not Nanobubbles: That is the Question. *Langmuir* **36**, 1699–1708 (2020).
22. Del Rosso, L. *et al.* Refined Structure of Metastable Ice XVII from Neutron Diffraction Measurements. *J. Phys. Chem. C* **120**, 26955–26959 (2016).
23. Jaramillo-Granada, A. M., Reyes-Figueroa, A. D. & Ruiz-Suárez, J. C. Xenon and Krypton Dissolved in Water Form Nanoblobs: No Evidence for Nanobubbles. *Phys. Rev. Lett.* **129**, 094501 (2022).
24. Takahashi, M., Shirai, Y. & Sugawa, S. Nanoshell Formation at the Electrically Charged Gas–Water Interface of Collapsing Microbubbles: Insights from Atomic Force Microscopy Imaging. *J. Phys. Chem. Lett.* **15**, 220–225 (2024).
25. Clark, G. N. I., Cappa, C. D., Smith, J. D., Saykally, R. J. & Head-Gordon, T. The structure of ambient water. *Molecular Physics* **108**, 1415–1433 (2010).
26. Ball, P. Water — an enduring mystery. *Nature* **452**, 291–292 (2008).
27. Kontogeorgis, G. M. *et al.* Water structure, properties and some applications – A review. *Chemical Thermodynamics and Thermal Analysis* **6**, 100053 (2022).
28. Finney, J. L. The structure of water: A historical perspective. *The Journal of Chemical Physics* **160**, 060901 (2024).
29. The Bakerian Lecture, 1962 The structure of liquids. *Proc. R. Soc. Lond. A* **280**, 299–322 (1964).
30. Stillinger, F. H. Water Revisited. *Science* **209**, 451–457 (1980).
31. Speedy, R. J. Self-replicating structures in water. *J. Phys. Chem.* **88**, 3364–3373 (1984).
32. *Supercooled Liquids: Advances and Novel Applications*. vol. 676 (American Chemical Society, 1997).
33. Yokoyama, H., Kannami, M. & Kanno, H. Existence of clathrate-like structures in supercooled water: X-ray diffraction evidence. *Chemical Physics Letters* **463**, 99–102 (2008).
34. Pathak, H. *et al.* Intermediate range O–O correlations in supercooled water down to 235 K. *The Journal of Chemical Physics* **150**, 224506 (2019).
35. Russo, J. & Tanaka, H. Understanding water's anomalies with locally favoured structures. *Nat Commun* **5**, 3556 (2014).

36. Santra, B., DiStasio, R. A., Martelli, F. & Car, R. Local structure analysis in *ab initio* liquid water. *Molecular Physics* **113**, 2829–2841 (2015).
37. Sun, T., Lin, F.-H., Campbell, R. L., Allingham, J. S. & Davies, P. L. An Antifreeze Protein Folds with an Interior Network of More Than 400 Semi-Clathrate Waters. *Science* **343**, 795–798 (2014).
38. Teeter, M. M. Water structure of a hydrophobic protein at atomic resolution: Pentagon rings of water molecules in crystals of crambin. *Proc. Natl. Acad. Sci. U.S.A.* **81**, 6014–6018 (1984).
39. Mpemba, E. B. & Osborne, D. G. Cool? *Phys. Educ.* **4**, 172–175 (1969).
40. Wojciechowski, B., Owczarek, I. & Bednarz, G. Freezing of aqueous solutions containing gases. *Cryst. Res. Technol.* **23**, 843–848 (1988).
41. Oishi-Tomiyasu, R. Robust powder auto-indexing using many peaks. *J Appl Crystallogr* **47**, 593–598 (2014).
42. Perdew, J. P., Burke, K. & Ernzerhof, M. Generalized Gradient Approximation Made Simple. *Phys. Rev. Lett.* **77**, 3865–3868 (1996).
43. Grimme, S. Semiempirical GGA-type density functional constructed with a long-range dispersion correction. *J Comput Chem* **27**, 1787–1799 (2006).
44. Ozaki, T. Variationally optimized atomic orbitals for large-scale electronic structures. *Phys. Rev. B* **67**, 155108 (2003).
45. Momma, K. & Izumi, F. *VESTA 3* for three-dimensional visualization of crystal, volumetric and morphology data. *J Appl Crystallogr* **44**, 1272–1276 (2011).
46. Seto, Y. & Ohtsuka, M. ReciPro: free and open-source multipurpose crystallographic software integrating a crystal model database and viewer, diffraction and microscopy simulators, and diffraction data analysis tools. *J Appl Crystallogr* **55**, 397–410 (2022).

**Acknowledgments:**

We thank Academia Sinica and the Ministry of Science and Technology (MOST 106-2112-M-001-025-MY3 and MOST 109-2112-M-001-048-MY3) of Taiwan, the National Science and Technology Council (NSTC 112-2112-M-001-047 and NSTC 113-2112-M-001-015) and Academia Sinica for supporting this study. The authors also acknowledge the technical support received from I.-H. Chen in the Advanced Materials Characterization Lab, Institute of Atomic & Molecular Sciences, Academia Sinica.

# Supplementary Materials

# Crystalline Water Structure in Room-Temperature Clathrate State: Hydrogen-Bonded Pentagonal Rings

Ching-Hsiu Chen[1], Wei-Hao Hsu[1], Ryoko Oishi-Tomiyasu[2], Chi-Cheng Lee[3], Ming-Wen Chu[4], and Ing-Shouh Hwang[1*]

[1] Institute of Physics, Academia Sinica, Nankang, Taipei 115, Taiwan, R.O.C.

[2] Institute of Mathematics for Industry, Kyushu University, Fukuoka, 819-0395, Japan

[3] Department of Physics, Tamkang University, New Taipei 251301, Taiwan

[4] Center for Condensed Matter Sciences, National Taiwan University, Taipei 10617, Taiwan

*Corresponding author. ishwang@phys.sinica.edu.tw (I.-S. Hwang)

**1. Supplementary Notes**

**2. Supplementary Figures**

**3. Supplementary Table**

**4. References**

# Supplementary Notes

## 1. Derivation of the lattice parameters and indexing the IZEDPs

Fig. S8 outlines how the indexing was performed. The code was implemented in C++, and the computation was rapid.

As illustrated in Fig. S8, the input data were peak-searching results of the in-zone electron diffraction patterns (IZEDPs). In step 1, Conway's topograph[1] was used to obtain potential candidates for the unit-cell parameters from combinations of zonal reflections. A topograph is a tree with infinitely many edges that correspond to 2D integral vectors.

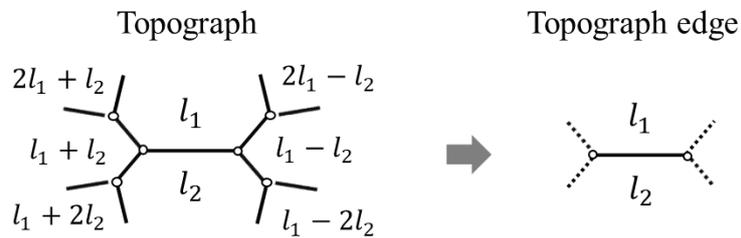

In this figure, every vector is a lax vector, that is, a $\pm l$ pair, which is simply denoted $l$, by ignoring the difference in the signs of vectors. For any lax vectors $l_1$ and $l_2$, the lax vector $l_3$ with $l_1 + l_2 + l_3 = 0$ is provided by $l_3 = l_1 + l_2$ or $l_1 - l_2$. Consequently, the edge corresponding to $l_1, l_2$ has two nodes associated with $\{l_1, l_2, -l_1 - l_2\}$ and $\{-l_1, l_2, l_1 - l_2\}$:

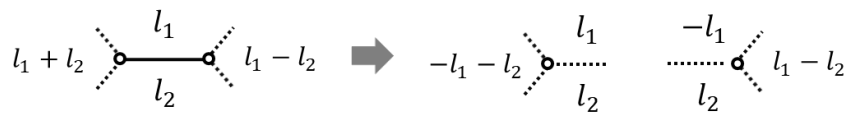

The lengths of the lattice vectors $l_1, l_2, l_1 \pm l_2$ satisfy the parallelogram law $[2(|l_1^*|^2 + |l_2^*|^2) = |l_1^* + l_2^*|^2 + |l_1^* - l_2^*|^2]$, which is also known as Ito's equation[2,3] in powder diffraction. In consideration of theorems on how forbidden reflections are distributed on topographs (the outlines are depicted in Figs. S9 and S10), topographs can be used to eliminate the adverse effects of forbidden reflections on the results of indexing.

The Bravais-lattice determination in step 2 involved lattice-basis reduction and centering processes, which were required to be error-stable, to avoid a failure in indexing that can occur when the observed parameters of the unit cell are reduced but its unknown true parameters are not. The method of lattice characters (Table 3.1.3.1) in International Table Vol. A[4] supports only unit-cell parameters with exact values. The method we used has been mathematically proven to be robust against large observation errors[5].

The algorithm of step 1 for IZEDPs, presented in the following, was developed as an analogue to the methods of the CONOGRAPH software for powder diffraction[6] and electron backscatter diffraction[7]. In the algorithm, (ii) of (2) addresses forbidden reflections and missing reflections for other reasons by estimating their coordinates from observed reflections.

**(Input)** image[$i$]: TEM diffraction images ($i$ = 1, 2, 3).
**(Output)** Ans: list of candidate parameters for the unit cell
**(Preprocess)** Perform a peak search for each image[$i$]. Reflections with $d$-values $d_{i1}, \dots, d_{is_i}$ are indexed by 2D integral vectors $v_{i1}, \dots, v_{is_i}$.

(1) **Preparation:** Let $b_{i1}$ and $b_{i2}$ be the three-dimensional vectors in the reciprocal space such that $l_{ij} = (b_{i1} \ \ b_{i2})v_{ij}$ provides the reflection corresponding to $v_{ij}$ in image[$i$]. In this situation, the length of $l_{ij}$ is given by the $d$-values $d_{ij}$:

$$\left|l_{ij}\right|^2 = v_{ij}^T \begin{pmatrix} b_{i1}^T \\ b_{i2}^T \end{pmatrix} (b_{i1} \ \ b_{i2}) v_{ij} = (1/d_{ij})^2.$$

Estimate the entries of the $2 \times 2$ symmetric matrix $S_2[i] = \begin{pmatrix} b_{i1}^T \\ b_{i2}^T \end{pmatrix}(b_{i1} \ \ b_{i2})$ by using the $d$-values $d_{i1}, \dots, d_{is_i}$.

(2) **Construction of topograph nodes:** For each distinct pair $v_1, v_2$ of $v_{ij}, \dots, v_{is_i}$, complete (i) and (ii). Using all the lax vectors in the nodes obtained in (i) and (ii), make a list A[$i$] ($i$ =1, 2, 3) of pairs of a lax vector $v$ and its corresponding $d$-value.
 (i) If any of $v_{ij}, \dots, v_{is_i}$ is equal to $v_1 + v_2$ or $-v_1 - v_2$, we have the following node:

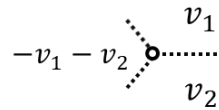

Additionally, if one of $v_{ij}, \dots, v_{is_i}$ is equal to $v_1 - v_2$ or $-v_1 + v_2$, we have the following node:

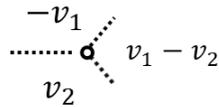

(ii) If one of $v_{ij}, \ldots, v_{is_i}$ is equal to $(v_1 + v_2)/2$ or $-(v_1 + v_2)/2$ and none of them are equal to $\pm(v_1 - v_2)/2$, or if one of $v_{ij}, \ldots, v_{is_i}$ is equal to $(v_1 - v_2)/2$ or $-(v_1 - v_2)/2$ and none of them are equal to $\pm(v_1 + v_2)/2$, we have the following two nodes:

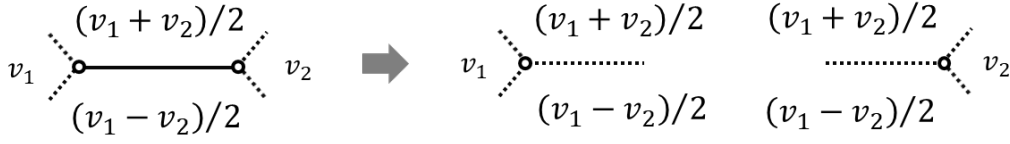

In this case, neither of the lax vectors $(v_1 + v_2)/2$ and $(v_1 - v_2)/2$ is observed, but the *d*-value can be calculated using Ito's equation.

(3) **Construction of unit-cell parameters:** For three pairs of entries $(u_i, d_i)$, $(v_i, \delta_i)$ of A[i] ($i = 1, 2, 3$) with *d*-values $d_1 \approx d_2$, $\delta_2 \approx \delta_3$, and $\delta_1 \approx d_3$ (i.e., equal within a margin of error), we assume that $u_i$ and $v_i$ satisfy either (i) or (ii):

(i) $(b_{11} \quad b_{12})u_1 = (b_{21} \quad b_{22})u_2$, $(b_{21} \quad b_{22})v_2 = (b_{31} \quad b_{32})v_3$, $(b_{11} \quad b_{12})v_1 = (b_{31} \quad b_{32})u_3$.

(ii) $(b_{11} \quad b_{12})u_1 = -(b_{21} \quad b_{22})u_2$, $(b_{21} \quad b_{22})v_2 = -(b_{31} \quad b_{32})v_3$, $(b_{11} \quad b_{12})v_1 = -(b_{31} \quad b_{32})u_3$.

In each case, we have the equations (i) $BM_+ = 0$ and (ii) $BM_- = 0$ with regard to the following matrices:

$$B = (b_{11} \quad b_{12} \quad b_{21} \quad b_{22} \quad b_{31} \quad b_{32}), \quad M_\pm = \begin{pmatrix} u_1 & 0 & v_1 \\ \pm u_2 & v_2 & 0 \\ 0 & \pm v_3 & \pm u_3 \end{pmatrix}$$

Calculate the Smith normal form of $M_+$ and $M_-$:

$$P_\pm M_\pm Q_\pm = \begin{pmatrix} d_1 & 0 & 0 \\ 0 & d_2 & 0 \\ 0 & 0 & d_3 \\ 0 & 0 & 0 \\ 0 & 0 & 0 \\ 0 & 0 & 0 \end{pmatrix},$$

$P_\pm, Q_\pm$: Square matrix with integral entries and determinant of $\pm 1$.

In each case, the reciprocal lattice is generated from the following $b_1, b_2, b_3$:

$$BP_\pm^{-1} = \begin{pmatrix} 0 & 0 & 0 \\ 0 & 0 & 0 & b_1 & b_2 & b_3 \\ 0 & 0 & 0 \end{pmatrix}.$$

Letting $P_{ij}^{(\pm)}$ be the $3 \times 2$ submatrices of $P_\pm = \begin{pmatrix} P_{11}^{(\pm)} & P_{12}^{(\pm)} & P_{13}^{(\pm)} \\ P_{21}^{(\pm)} & P_{22}^{(\pm)} & P_{23}^{(\pm)} \end{pmatrix}$,

$$B = \begin{pmatrix} b_1 & b_2 & b_3 \end{pmatrix} \begin{pmatrix} P_{21}^{(\pm)} & P_{22}^{(\pm)} & P_{23}^{(\pm)} \end{pmatrix}.$$

The parameters $S = (b_i \cdot b_j)$ of the reciprocal lattice are obtained by solving the overdetermined system of linear equations on the entries of $S$.
$$P_{2i}^{(\pm)\,T} S P_{2i}^{(\pm)} = S_2[i] \quad (i = 1, 2, 3).$$

Because $S$ is a $3 \times 3$ symmetric matrix, the number of independent variables is 6, and the number of equations is 9. For each $S$ obtained using the aforementioned procedures, the unit-cell parameters $a, b, c, \alpha, \beta$, and $\gamma$ can be calculated using the following formula:

$$S^{-1} = \begin{pmatrix} a^2 & ab\cos\gamma & ac\cos\beta \\ ab\cos\gamma & b^2 & bc\cos\alpha \\ ac\cos\beta & bc\cos\alpha & c^2 \end{pmatrix}.$$

## 2. Low-symmetry P1 space group for CS-RT

Crystals with a hexagonal unit cell typically have symmetry higher than P1. We investigated such a possibility for the crystalline water structure in mesoscopic clathrate hydrate structures. The condition with forbidden reflections of 00$l$, where $l$ is an odd number, limits the possible space groups to P6_3 (#173), P6_3/m (#176), and P6_322 (#182). For these space groups, however, we could not identify any crystalline water structure for which all O–O distances were within the range 2.60–3.6 Å. Thus, no crystalline water structures with symmetry higher than P1 fit the experimental IZEDPs. We speculated that entropy may favor the low-symmetry P1 space group at RT, which is higher than the typical temperatures for formation of water ices. Hydrogen bonds are weak (relative to typical chemical bonds), and thus, the bond lengths and related angles in CS-RT may vary to a certain degree because of the relatively large thermal energies of the water molecules. The crystalline water structures may have a triclinic unit cell that slightly deviates from the hexagonal lattice momentarily and locally but remains in a hexagonal lattice in general. This would mean that water molecules may not be limited to thermal vibrations around the lattice sites of the perfect hexagonal unit cell. Compared with that in other highly symmetric hexagonal structures, the higher degree of freedom in the thermal movements of water molecules in the low-symmetry CS-RT crystal means higher entropy, which contributes to lower free energy.

## 3. Type α IZEDP

On the basis of the CS-RT structure, identifying one IZEDP of Type α within our sample tilting range would be easy if the *c*-axis were roughly perpendicular to the electron beam direction. A Type α IZEDP is an IZEDP along the *a*-axis ([100]), *b*-axis ([010]), or [110]. These orientations are perpendicular to the *c*-axis, [001], indicating six IZEDPs are separated by approximately 60°. Fig. S11 presents the simulated structures and IZEDPs of Type α at six orientations. Slight differences were noted in the structures and IZEDPs for the six orientations. In particular, the two diffraction spots at approximately 7.1 Å (001 and $00\bar{1}$ reflections) had different intensities. This is because the derived structure has P1 symmetry, which is not perfect six-fold symmetry. To achieve the same intensity for the two reflections (001 and $00\bar{1}$), the sample must be tilted at a small angle (0.01°–0.06°) around an axis perpendicular to the **c**-axis, which varies for the six different rotations.

## 4. Effect of GLCs on the clathrate state

Researchers have reported on high pressure inside GLCs. Whether mesoscopic clathrate structures exist because of the high pressure and/or the nanometer confinement in GLCs remains unclear. Molecules confined between graphene layers have been proposed to potentially experience van der Waals pressure, which can be estimated as $P_W \cong E_W/d$, where $E_W \approx$ 20–30 meVÅ$^{-2}$ is the adhesion energy and $d$ is the thickness of the material sandwiched between the graphene layers[8]. The pressure can be as high as approximately 1 GPa when the confined thickness is smaller than 1 nm, but the pressure decreases with increasing thicknesss[8–10]. Force indentation experiments performed using atomic force microscopy indicated that $P_W \approx$ 3 MPa when $d \approx$ 26 nm.[11] The thickness of water pockets containing mesoscopic clathrate structures was estimated through electron loss energy spectroscopy to be 20–100 nm[12], and the pressure was thus estimated to be smaller than 5 MPa. This pressure is far from sufficient to drive the phase transition of liquid water into water ices. Water ices VI, VII, and X are stable at room temperature but form at a pressure of approximately 1 GPa or higher, and their structures are not compatible with the diffraction patterns presented herein. In addition, various TEM studies of aqueous solutions in GLCs have been conducted[13–16]; water has been found to remain in the liquid state, and mesoscopic clathrate structures have been discovered only in gas-supersaturated water[12,17]. These observations indicate that the pressure inside GLCs thicker than approximately 20 nm is insufficient to drive the crystallization of water. Although the exact pressure inside a GLC is unknown, the confinement of gas-supersaturated water inside GLCs may have effects. First, the shapes of mesoscopic clathrate structures in bulk water may be approximately spherical, but they become flatter (thickness ≈ 1/10 lateral size) and distorted due to nonuniform confinement by graphene layers. Second, the pressure in GLCs may vary from one water pocket to another due to variation in water thickness and the shape of the water pocket. The pressure variation may also affect the variation of the d-spacings; however, this must be further investigated.

**Supplementary Figures**

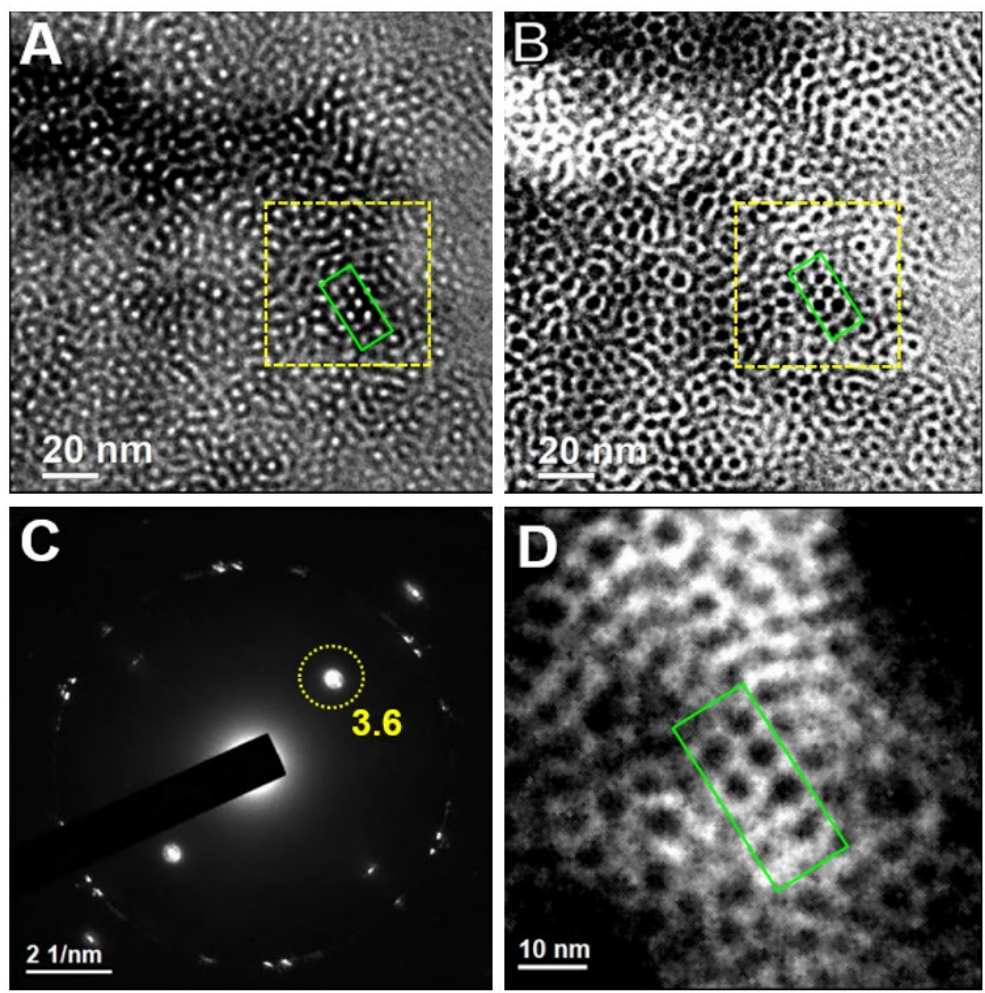

**Fig. S1** TEM bright-field and dark-field imaging of a mesoscopic clathrate structure in an ethanol-water (EW) mixture sandwiched in a graphene liquid cell. (A,B) Bright-field images acquired at underfocus and overfocus, respectively. (C) SAED of the clathrate structure. Two strong diffraction spots with a d-spacing of 3.6 Å are evident. A ring of first-order diffraction spots of the graphene lattice (d-spacing of 2.14 Å) is visible outside the strong diffraction spots. (D) Dark-field image recorded using the diffraction spot outlined with a dotted yellow circle in (C). Only the region outlined by a yellow square in (A,B) is shown. The green rectangle outlines a small region for comparison between the images in (A, B) and (D). Note that the crystals that contribute to the diffracted beam appear bright.

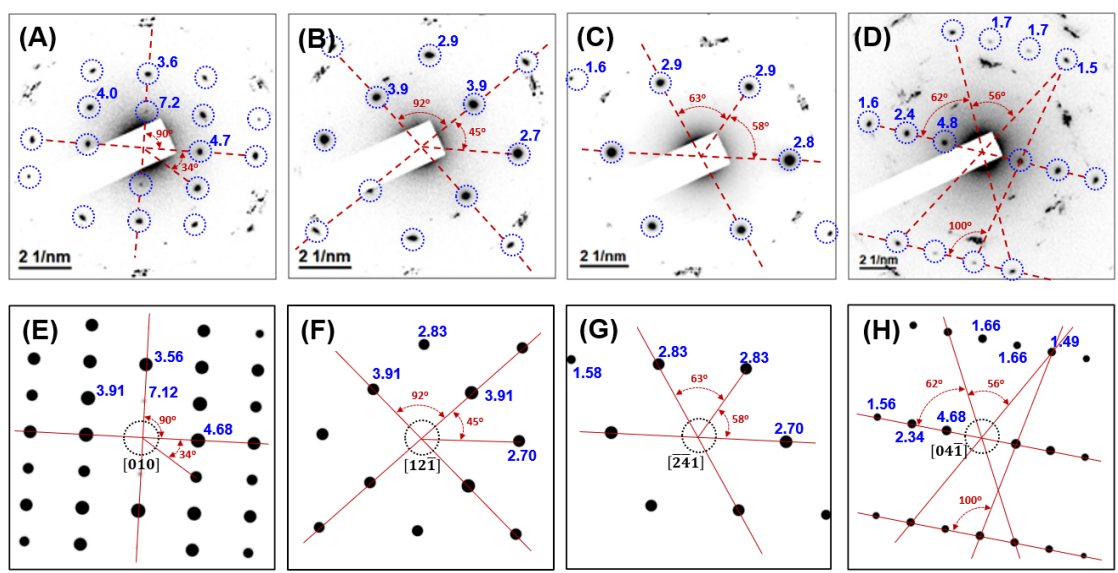

**Fig. S2** Detailed comparison of experimental and simulated IZEDPs. (A–D) Experimental IZEDPS, which are the same as in Fig. 1A–D. The number next to some diffraction spots indicates the d-spacing (unit Å). (E–H) Simulated IZEDPS, which are the same as those in Fig. 1E–H. The related d-spacing and angles obtained in the experiments versus simulations match favorably.

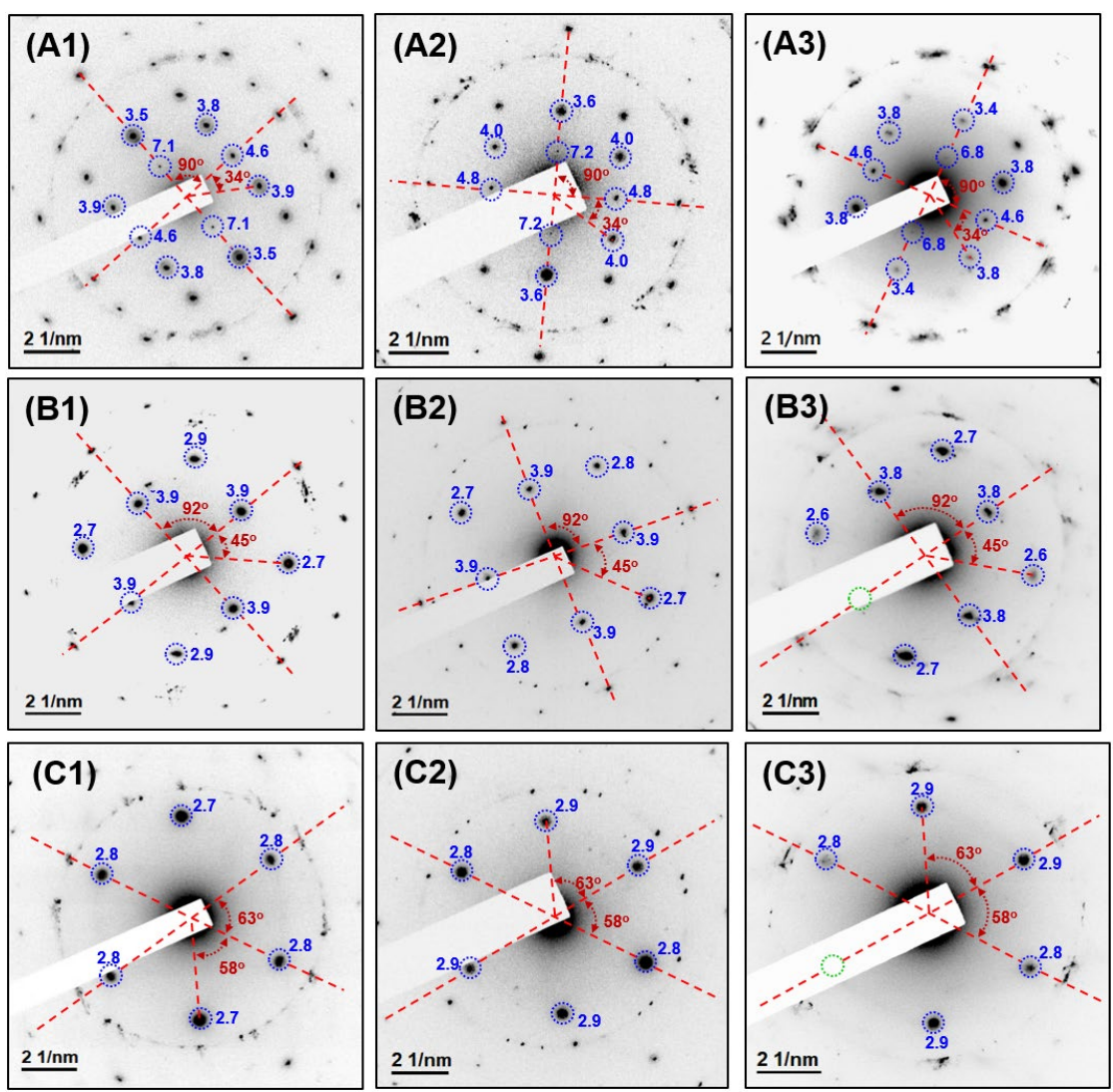

**Fig. S3** IZEDPs of Type α (A1–A3), Type β (B1–B3), and Type γ (C1–C3) recorded for different mesoscopic clathrate structures. The number next to some diffraction spots indicates the d-spacing (unit Å).

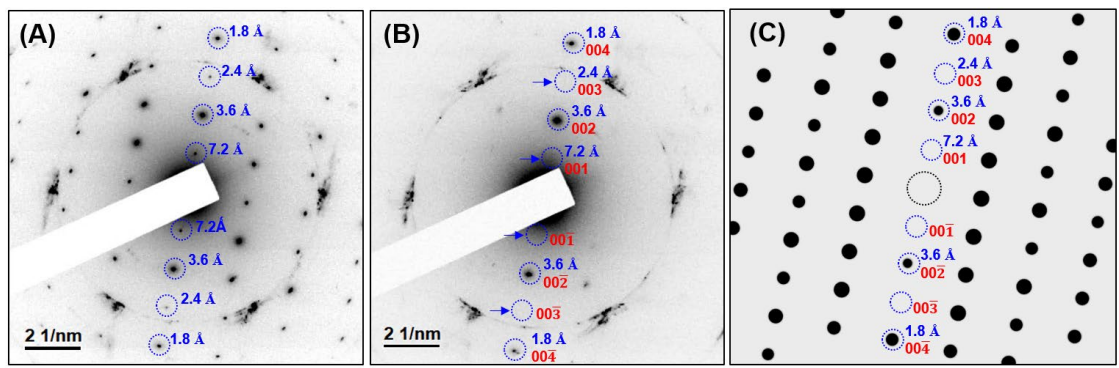

**Fig. S4** Forbidden reflections for 00*l*, where *l* is an odd number. (A) IZEDPs of Type α recorded for a mesoscopic clathrate structure. (B) SAED pattern of the mesoscopic clathrate structure after the sample is rotated 7.3° around the *c*-axis. The reflections 00*l*, where *l* is an odd number, disappear. Arrows indicate four forbidden reflections, 001, 00$\bar{1}$, 003, and 00$\bar{3}$. (C) Simulated kinematic IZEDPs of Type α. The reflections 00*l*, where *l* is an even number, are allowed.

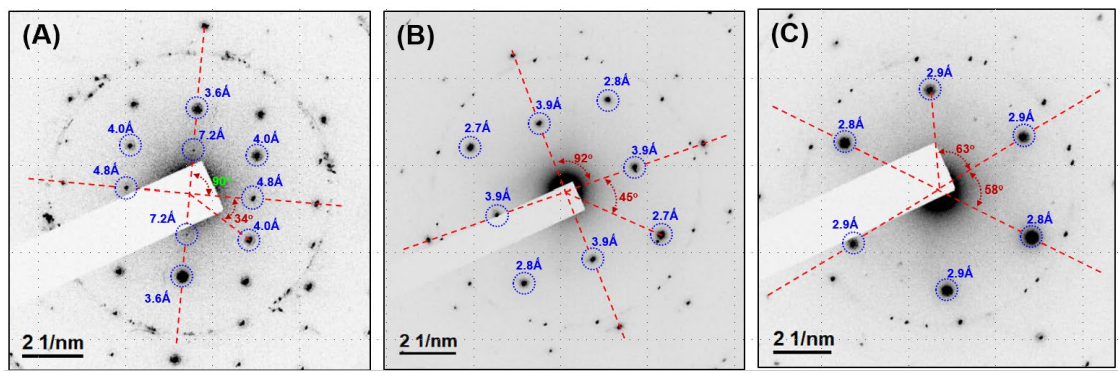

**Fig. S5** Experimental IZEDPs recorded on Clathrate 2, which is a clathrate structure in N$_2$-supersaturated water sandwiched in a GLC. (A–C) Types α, β, and γ, respectively.

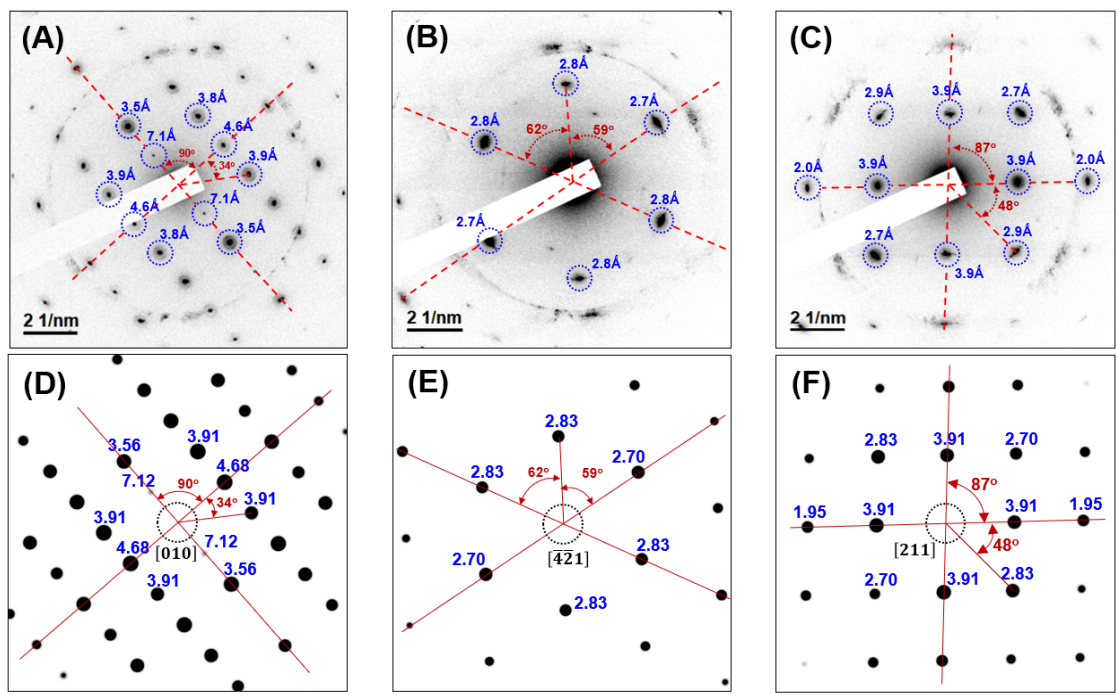

**Fig. S6** Experimental and simulated IZEDPs of Clathrate 3, which is a clathrate structure in an EW mixture sandwiched in a GLC. (A–C) Types α, γ, and ε, respectively. (D–F) Simulated IZEDPs of Types α, γ, and ε, respectively.

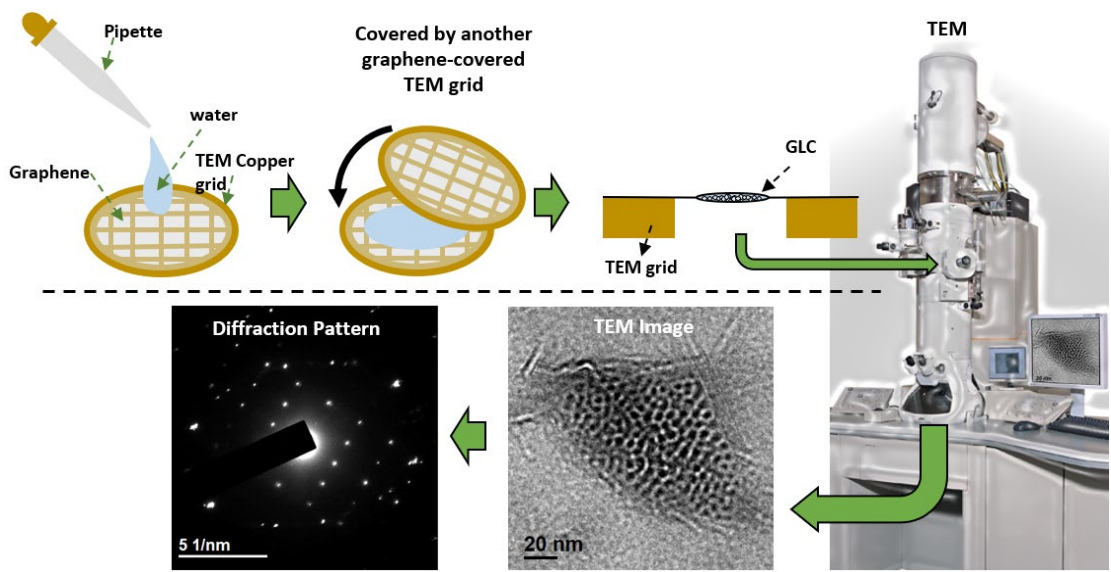

**Fig. S7** Schematic of procedure for preparing water sandwiched in graphene liquid cells and TEM characterization of mesoscopic clathrate structures.

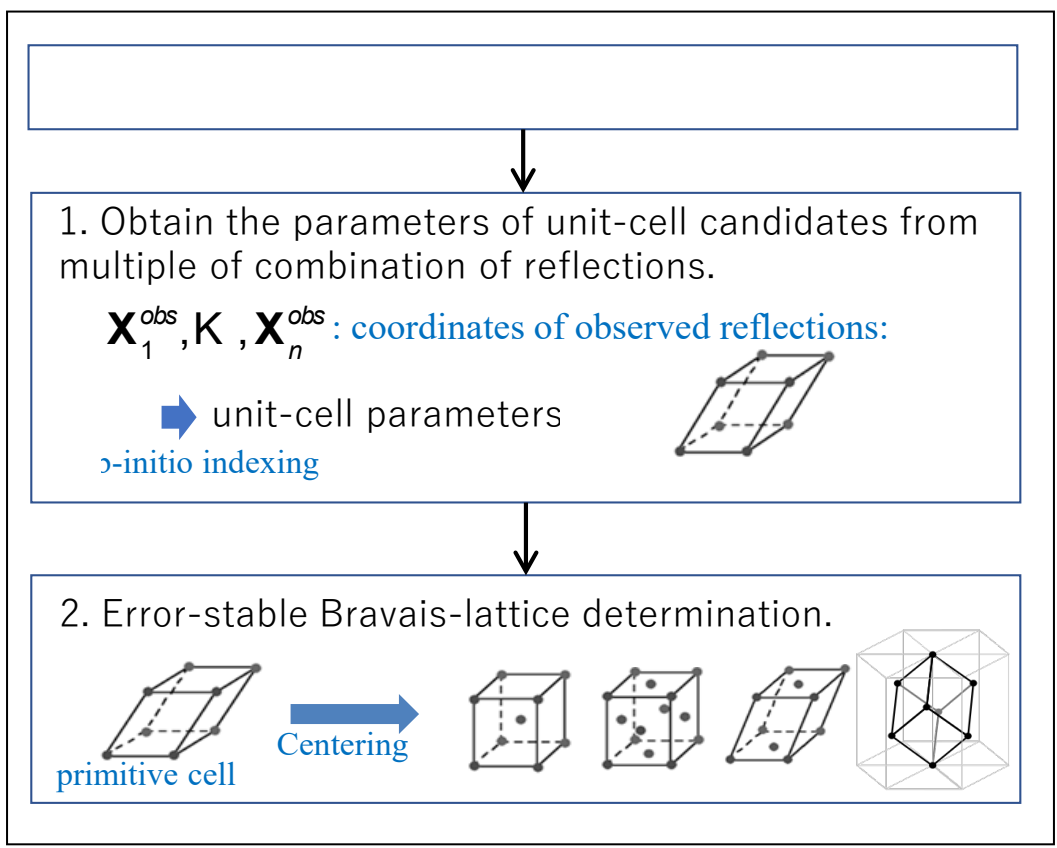

**Fig. S8** Flowchart of ab-initio indexing.

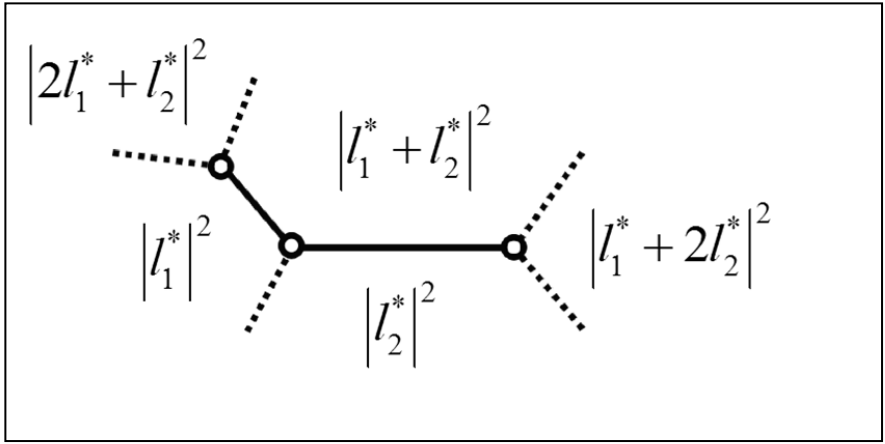

**Fig. S9** Location of reciprocal lattice vectors $l_1^*, l_2^*, l_1^* + 2l_2^*, 2l_1^* + l_2^*$ in the topograph. None of them are forbidden reflections in Th.2 [18].

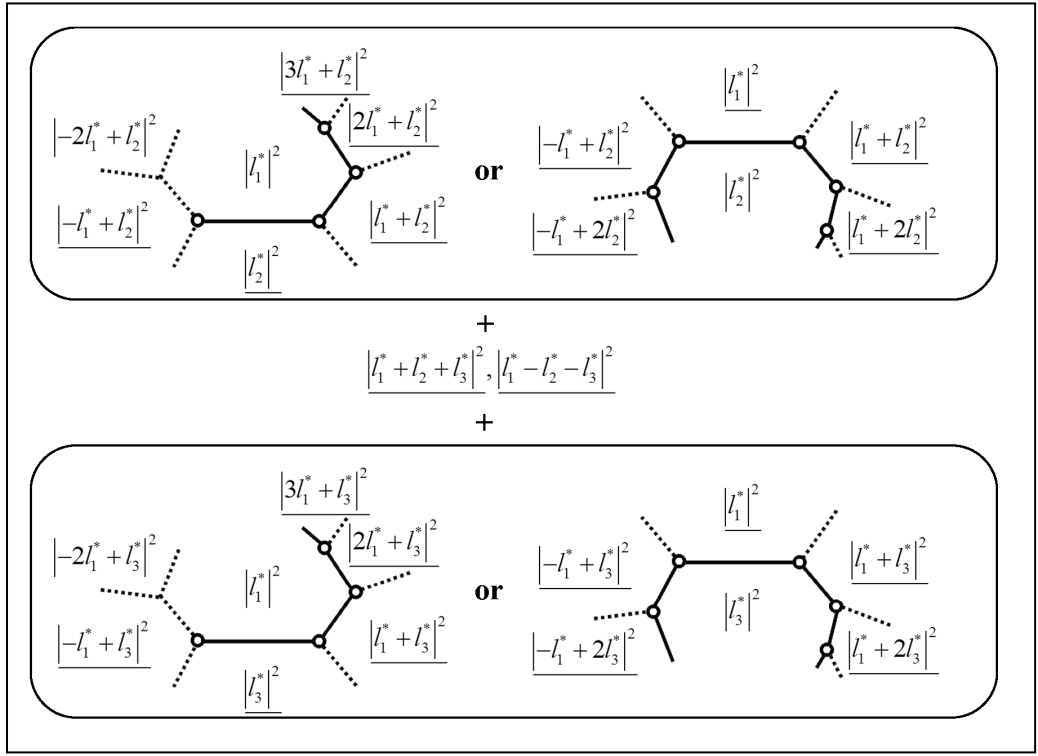

**Fig. S10** Location of reciprocal lattice vectors in the topograph. None of underlined vectors are forbidden reflections in Th.4 [14].

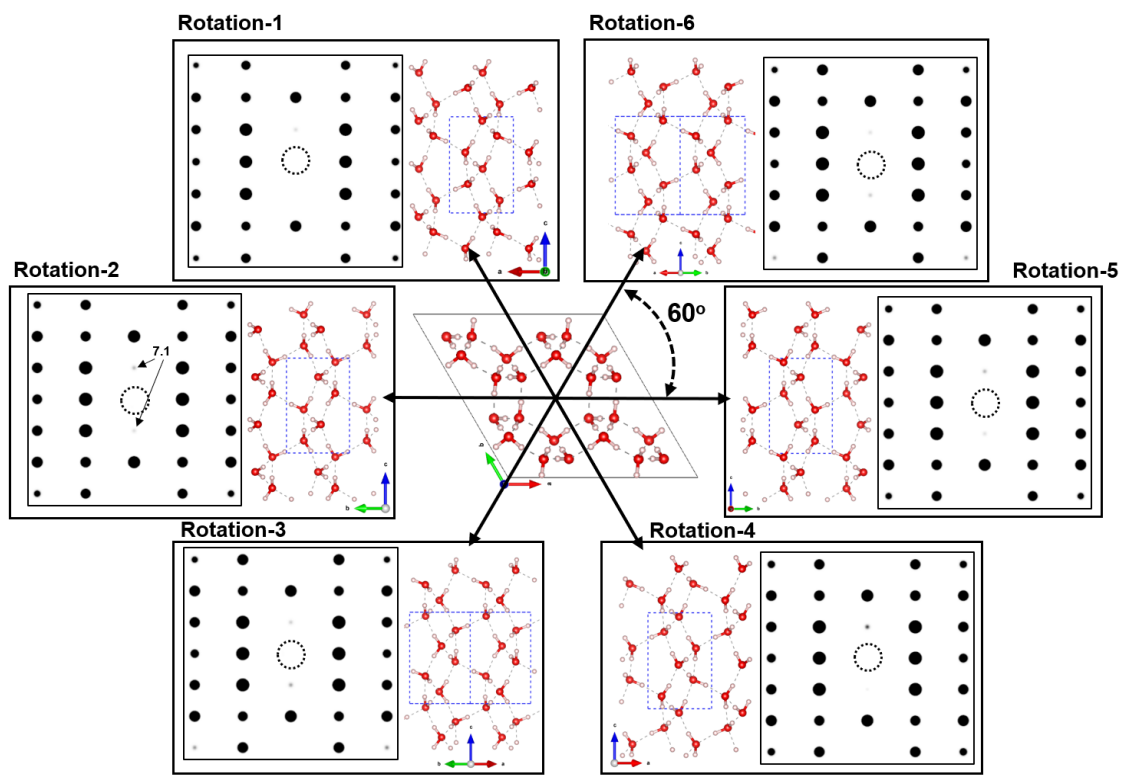

**Fig. S11** Six quasisymmetric Type α diffraction patterns with the *c*-axis as the rotation axis and the related atomic models. The dashed blue rectangle represents a single unit cell.

# Supplementary Table

**Table S1** Experimental relative angles between different IZEDPs recorded on Clathrate 3 are compared with the simulated angles between three types of IZEDPs.

|        | Clathrate 3 | | Simulation | |
| --- | --- | --- | --- | --- |
|        | **Type-A** | **Type-C** | **Type-A** | **Type-C** |
| **Type-C** | 35.0° | --- | 35.9° | --- |
| **Type-E** | 45.5° | 57.1° | 46.4° | 58.1° |

# References

1. Conway, J. H. & Fung, F. Y. C. *The Sensual Quadratic Form*. vol. 26 (Mathematical Association of America, 1997).
2. Ito, T. A General Powder X-Ray Photography. *Nature* **164**, 755–756 (1949).
3. De Wolff, P. M. On the determination of unit-cell dimensions from powder diffraction patterns. *Acta Cryst* **10**, 590–595 (1957).
4. *International Tables for Crystallography: Space-group symmetry*. vol. A (International Union of Crystallography, 2016).
5. Oishi-Tomiyasu, R. Rapid Bravais-lattice determination algorithm for lattice parameters containing large observation errors. *Acta Crystallogr A Found Crystallogr* **68**, 525–535 (2012).
6. Oishi-Tomiyasu, R. Robust powder auto-indexing using many peaks. *J Appl Crystallogr* **47**, 593–598 (2014).
7. Oishi-Tomiyasu, R., Tanaka, T. & Nakagawa, J. Distribution rules of systematic absences and generalized de Wolff figures of merit applied to electron backscatter diffraction *ab initio* indexing. *J Appl Crystallogr* **54**, 624–635 (2021).
8. Algara-Siller, G. *et al.* Square ice in graphene nanocapillaries. *Nature* **519**, 443–445 (2015).
9. Vasu, K. S. *et al.* Van der Waals pressure and its effect on trapped interlayer molecules. *Nat Commun* **7**, 12168 (2016).
10. Gao, Z., Giovambattista, N. & Sahin, O. Phase Diagram of Water Confined by Graphene. *Sci Rep* **8**, 6228 (2018).
11. Khestanova, E., Guinea, F., Fumagalli, L., Geim, A. K. & Grigorieva, I. V. Universal shape and pressure inside bubbles appearing in van der Waals heterostructures. *Nat Commun* **7**, 12587 (2016).
12. Hsu, W.-H. & Hwang, I.-S. Investigating states of gas in water encapsulated between graphene layers. *Chem. Sci.* **12**, 2635–2645 (2021).
13. Yuk, J. M. *et al.* High-Resolution EM of Colloidal Nanocrystal Growth Using Graphene Liquid Cells. *Science* **336**, 61–64 (2012).
14. Wang, C., Qiao, Q., Shokuhfar, T. & Klie, R. F. High-Resolution Electron Microscopy and Spectroscopy of Ferritin in Biocompatible Graphene Liquid Cells and Graphene Sandwiches. *Advanced Materials* **26**, 3410–3414 (2014).
15. Kim, B. H. *et al.* Liquid-Phase Transmission Electron Microscopy for Studying Colloidal Inorganic Nanoparticles. *Advanced Materials* **30**, 1703316 (2018).
16. Wu, H., Friedrich, H., Patterson, J. P., Sommerdijk, N. A. J. M. & De Jonge, N. Liquid-Phase Electron Microscopy for Soft Matter Science and Biology.

*Advanced Materials* **32**, 2001582 (2020).

17. Hsu, W.-H. *et al.* Observation of mesoscopic clathrate structures in ethanol-water mixtures. *Journal of Molecular Liquids* **366**, 120299 (2022).
18. Oishi-Tomiyasu, R. Distribution rules of systematic absences on the Conway topograph and their application to powder auto-indexing. *Acta Crystallogr A Found Crystallogr* **69**, 603–610 (2013).